\documentclass[%
superscriptaddress,
preprint,
nofootinbib,
 amsmath,amssymb,
aps,
pra,
endfloats,
floatfix
]{revtex4-2}

\usepackage[utf8]{inputenc}
\usepackage[T1]{fontenc}
\usepackage{graphicx}
\usepackage{dcolumn}
\usepackage{bm}
\usepackage{siunitx} 
\usepackage[french, main =english]{babel}
\usepackage[hidelinks]{hyperref}
\usepackage[mathlines]{lineno}
\usepackage[version=4]{mhchem} 
\usepackage{miller}
\usepackage{booktabs}
\graphicspath{figures}

\usepackage{hyphenat}
\begin{document}

\title{
Accurate prediction of optical transitions in epitaxial InGaAs/InAlAs asymmetric coupled quantum well structures
}

\author{Konstantinos Pantzas}
\affiliation{%
 Centre de Nanosciences et de Nanotechnologies, CNRS - Université Paris-Saclay, 91120, Palaiseau, France
}%

\author{Virginie Trinité}
\affiliation{III-V Lab, 91120, Palaiseau, France}

\author{Angela Vasanelli}
\affiliation{Laboratoire de Physique de l’École Normale Supérieure,
ENS, Université, PSL, CNRS, Sorbonne Université,
Université Paris Cité, F-75005 Paris, France}

\author{Carlo Sirtori}
\affiliation{Laboratoire de Physique de l’École Normale Supérieure,
ENS, Université, PSL, CNRS, Sorbonne Université,
Université Paris Cité, F-75005 Paris, France}

\author{Grégoire Beaudoin}
\affiliation{%
 Centre de Nanosciences et de Nanotechnologies, CNRS - Université Paris-Saclay, 91120, Palaiseau, France
}%

\author{Isabelle Sagnes}
\affiliation{%
 Centre de Nanosciences et de Nanotechnologies, CNRS - Université Paris-Saclay, 91120, Palaiseau, France
}%

\author{Jean-Luc Reverchon}
\affiliation{III-V Lab, 91120, Palaiseau, France}

\author{Gilles Patriarche}
\affiliation{%
 Centre de Nanosciences et de Nanotechnologies, CNRS - Université Paris-Saclay, 91120, Palaiseau, France
}%
\begin{abstract}

Atomically-resolved Z-contrast and strain mappings are used to extract a model of the composition of an InGaAs/InAlAs asymmetric coupled quantum-well structure grown on InP using metal-organic vapor phase epitaxy. The model accounts for grading across the multiple alloy interfaces. The model is used to compute intersubband absorption in the structure. The simulation accurately predicts the experimental absorption spectrum of the structure within only a few \si{\milli\electronvolt}, an almost ten-fold improvement over simulations using a square-band profile with nominal alloy compositions, and a significant step forward in accurate and predictive simulations of the optical properties epitaxial heterostructures for emission, modulation and detection in mid-infrared.
\end{abstract}

\maketitle

\section{Introduction}\label{sec:introduction}

Optoelectronic devices operating in the mid-infrared domain are required for several applications, including night vision, free-space communications \cite{martini_high-speed_2001}, light detection and ranging (LIDAR) \cite{diaz_active_2016}, spectroscopy \cite{villares_dual-comb_2014}, and observational astronomy \cite{hale_berkeley_2000}. Unipolar quantum devices, i.e. quantum cascade lasers (QCLs), detectors (QCDs) and quantum-well infrared  detectors (QWIPs), as well as interband cascade lasers (ICLs) offer unique possibilities for room-temperature and high-frequency operation. These devices are now all reaching a level of maturity sufficient for the development of mid-infrared optoelectronic systems, with recent examples include systems for free-space communication combining a QCL - modulated either directly  or using an external modulator - and a high-frequency quantum-well detector \cite{dely10GbitFree2022,saemian_ultra-sensitive_2024}. 

Further development of such systems hinges on a better understanding and modeling of the complex heterostructures that underpin the individual devices. Indeed, devices based on semiconductor superlattices rely on designs involving several hundred wells and barriers, with some designs calling for layers only a fraction of an atomic monolayer and even minute deviations in the epitaxial structure induce large shifts of the operating wavelength. In this context, fine-tuning a given device design to improve overall system performance is a time consuming and laborious process involving numerous cycles of epitaxy, processing and characterization.

One of the key issues is understanding, modeling, and controlling interfaces. In most semiconductor devices, individual layers in structures are sufficiently thick to consider interfaces ideally abrupt and model the potential in these crystals using square wells and barriers. Epitaxial interfaces, however, have a finite width, comparable or greater than  the width of certain layers in the design of unipolar devices, and, as a result, greatly impact their performance. Developments in characterization techniques recently made it possible to measure the characteristic interface length of epitaxial interfaces \cite{ashuachQuantificationAtomicIntermixing2013,patilGaAsBiGaAsMultiquantum2017,lunaCriticalRoleTwoDimensional2012,lunaInterfacePropertiesGa2009,lunaIndiumDistributionInterfaces2008,hulkoComparisonQuantumWell2008,prosaAtomProbeAnalysis2011,grangeAtomicScaleInsightsSemiconductor2020,mullerInterfacialChemistryInAs2012,wangSensitivityHeterointerfacesEmission2017}. 

These works highlight the importance of investigating and more accurately modeling the microstructure of epitaxial crystals used in unipolar devices. Nevertheless, the models proposed in these works exhibit a few shortcomings. First, most studies have been limited  to heterostructures where the characteristic length of interfaces is  smaller than the width of the thinnest layers. Furthermore, the studies often consider interfaces where only one element is exchanged - e.g. Ge/SiGe \cite{grangeAtomicScaleInsightsSemiconductor2020} or GaAs/AlGaAs \cite{lunaCriticalRoleTwoDimensional2012}.   However, several key structures in unipolar devices rely on alloys where more than one elements are exchanged across interfaces. Finally, the case of multiple interfaces spaced closely together and the ensuing influence on grading is not handled.

The present contribution aims to address these shortcomings. A generalized expression for compositional grading for arbitrary number of interfaces, spacings, and number of chemical elements in alloys is proposed. The use of the model is showcased in the design of asymmetric coupled quantum wells (ACQW) using  (In,Al,Ga)As alloys lattice-matched on InP. The model is calibrated on experimental data  obtained from quantified scanning transmission electron microscopy. It is then  used in numerical simulations and shown to accurately predict the experimental absorption spectra of the ACQW structures. 
\section{Models and Methods}\label{sec:methods}

Compositional grading at interfaces can be modeled with a variety of mathematical sigmoid functions. Two such functions have been used in the literature,  the logistic function~\cite{lunaCriticalRoleTwoDimensional2012,lunaIndiumDistributionInterfaces2008,lunaInterfacialIntermixingInAs2010,lunaInterfacePropertiesGa2009} and the error function \cite{grangeAtomicScaleInsightsSemiconductor2020}. Both functions were shown to yield a good agreement with experiment. The logistic function is, however, the simpler of the two and better suited to the generalization proposed here. Following Reference~\cite{lunaCriticalRoleTwoDimensional2012}, in a crystal for which the composition of different layers varies across an interface at position $z_n$, the variation due to grading at an interface is modeled using:
\begin{equation}
   \theta_n\left(z\right) = 1 - \frac{1}{1+e^{\left(z-z_n\right)/L_{n}}},
\label{eq:def_sigmoid} 
\end{equation}
where $L_{n}$ is the characteristic length of transition between the two layers of different compositions on either side of the interface at $z_n$. 
Now, consider a crystal consisting of $N$ layers, ordered along direction $z$, and made of various alloys. Let $M$ be the number of total elements in the crystal, and $c_{m,n}$ the target composition far from any interface of alloy element $m$ in layer $n$. The sum of the compositions over all alloy elements must obey:
\begin{equation}
    \sum_{m=0}^{M-1} c_{m,n} = 1
\end{equation}
at every point in the crystal. Taking grading into account, the composition of  element $m$, $\mathcal{C}_{m}(z)$, at each position $z$ is given by:
\begin{equation}
       \mathcal{C}_{m}\left(z\right) = c_{m,0}+\sum_{n=0}^{N-1}\left(c_{m,n+1}-c_{m,n}\right)\theta_{n}\left(z\right),
\label{eq:sumNL} 
\end{equation}
i.e. the sum of contributions from all layers in the crystal, weighted by $\theta_{n}$ that accounts for grading across interfaces. This expression meets the criteria set out above: it accounts for multiple layers, arbitrarily spaced and taking into account all elements in the structure. One can readily show that the sum over $m$ of all $\mathcal{C}_m$ also equals unity at all $z$ in the crystal. The parameters in this expression -  the positions of the interfaces $z_n$, the characteristic lengths $L_n$, and the compositions $c_{m,n}$ - are determined  by fitting the model to  experimental data from scanning transmission electron microscopy.

To calibrate the model, an ACQW structure was grown in a Veeco D180 Turbodisc MOVPE reactor (sample A). The structure consists of two InGaAs wells that are \SI{5.6}{\nano\meter} and \SI{2.5}{\nano\meter} thick, separated by a \SI{1.4}{\nano\meter} thick InAlAs barrier. This core structure is surrounded by \SI{25}{\nano\meter} thick InAlAs barriers. Trimethylgallium, trimethylaluminum, and trimethylindium were used as organometallic precursors to elementary gallium, aluminum, and indium, respectively, and arsine as a hydride precursor to elementary arsenic. All alloys are lattice-matched on InP. The thin InAlAs barrier was grown following the aluminum flow modulation scheme introduced in Reference~\cite{pantzasSubnanometricallyResolvedChemical2016} to compensate for the partial loss of aluminum in very thin barriers. 

A lamella was prepared from sample A using ion milling and thinning, carried out in a FEI SCIOS dual-beam FIB-SEM. It was then observed in an aberration-corrected FEI TITAN 200 TEM-STEM  operating at \SI{200}{\kilo\electronvolt}. The convergence half-angle of the probe was \SI{17.6}{\milli\radian} and the detection inner and outer half-angles for HAADF-STEM were \SI{69}{\milli\radian} and \SI{200}{\milli\radian}, respectively. The lamella was imaged along the \hkl<110> zone axis. Atomically-resolved mappings of composition and strain where obtained from the HAADF-STEM micrograph following the algorithm detailed in Reference~\cite{pantzasExperimentalQuantificationAtomicallyresolved2021}.

A separate sample was grown containing thirty repetitions of the ACQW structure, to carry out multi-pass absorption experiments (sample B). In this structure the \SI{5.6}{\nano\meter} InGaAs well was n-doped with Si, with a target concentration of \SI{1.5e18}{\per\centi\meter\cubed}. Furthermore, a semi-insulating InP substrate was used in sample B to avoid free-carrier absorption. The absorption measurement is carried out with a Fourier-Transform Infrared (FTIR) spectrometer using and a multi-pass absorption cavity formed by two gold mirrors placed on either side of the sample. The length of the cavity can be controlled to change the number of passes. The sample is inserted inside the cavity and positioned at Brewster angle, and the experiment is carried out in a dry nitrogen atmosphere. The light generated by a globar is filtered with a linear TM polarizer before being injected. The detector in this setup is a broadband DLaTGS. The baseline is normalized by dividing the sample transmission by that measured transmission on the semi-insulating InP substrate alone. 

Band-structure simulations were carried out using an in house code that calculates the band-structure in the envelope function approximation and takes into account Poisson redistribution of charges and non-parabolicity, as described in Reference~\cite{Sirtori1994}. The absorption is calculated through Fermi's golden rule, taking into account the difference in effective mass of each level as in Reference~\cite{terazzi_transport_2012}. The maximum absorption in this case is red-shifted with respect to the transitions energies between levels computed in the structure. For the sake of brevity though, the absorption maximum corresponding to transition between levels $E_i$ and $E_j$ is noted $E_{ij}$. Finally a depolarization shift is used to account for doping in the wells.

\section{Results and Discussion}\label{sec:rnd}

\begin{figure}
    \centering
    \includegraphics[width=\textwidth]{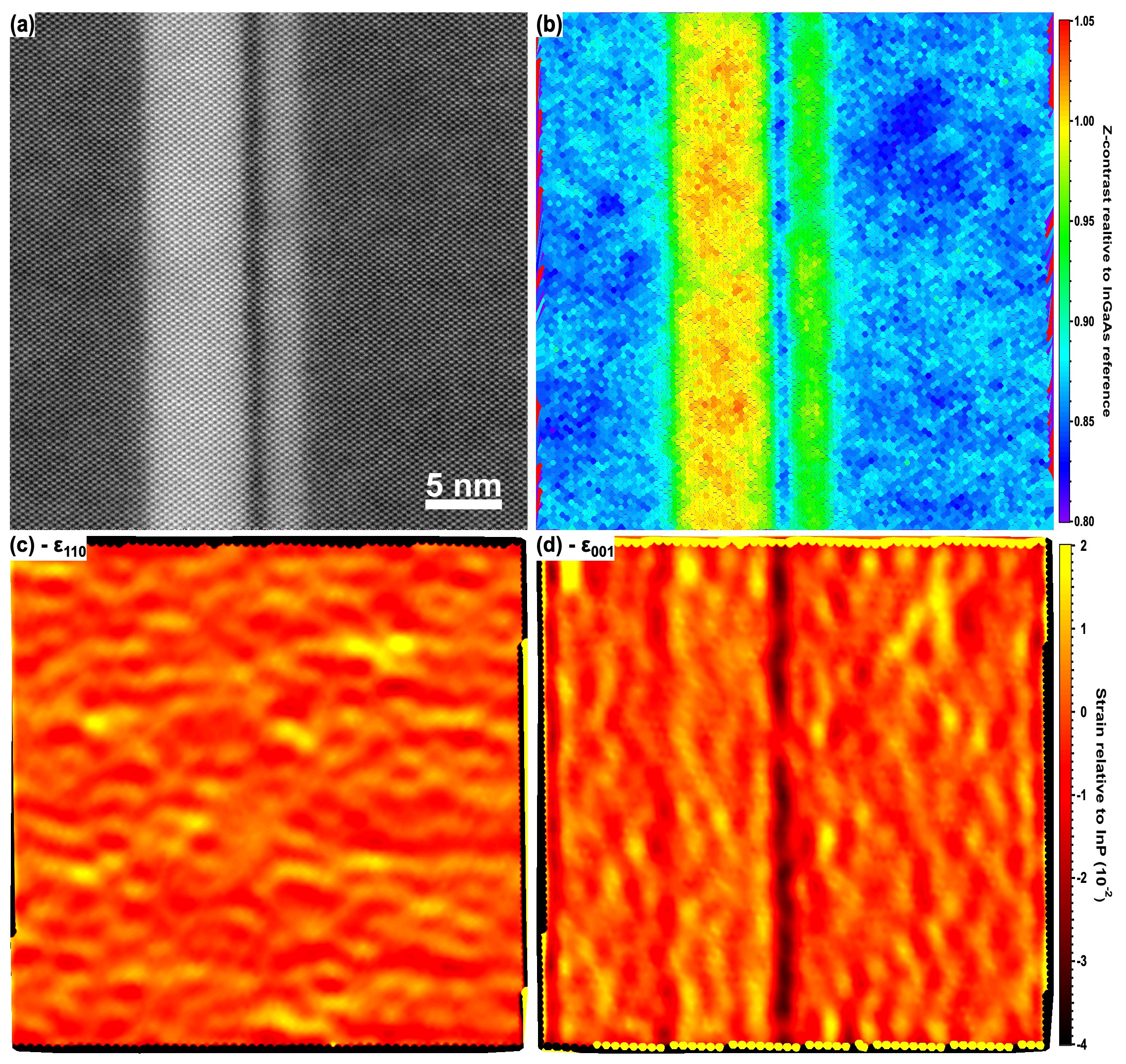}
    \caption{(a) HAADF-STEM micrograph of the asymmetric coupled quantum well structure (b) Corresponding atomically-resolved mapping of the aluminum concentration. (c) Displacement map along the in-plane \hkl<110> direction - no strain is observed, showing that the structure is pseudomorphically accommodated on the InP substrate (d) Displacement map along the \hkl<001> growth direction. The displacement deviates from the lattice parameter of InP for the InAlGaAs barrier indicating a departure from lattice-matched compositions of the InAlGaAs barrier.}
    \label{fig:acqw_mappings}
\end{figure}

\begin{figure}
    \centering
    \includegraphics[width=\textwidth]{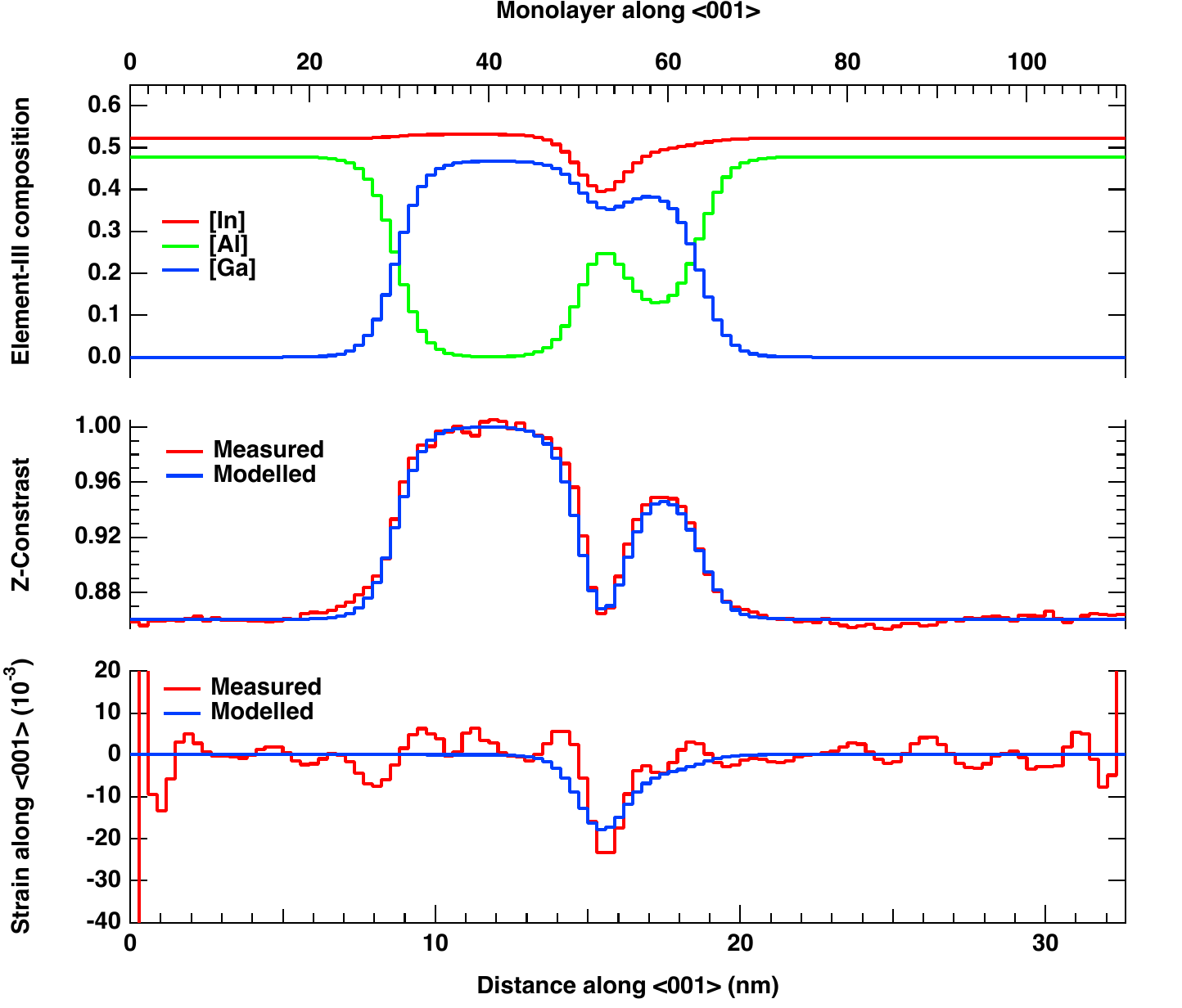}
    \caption{Composition profiles computed using the sigmoidal model to analyze the Z-contrast and displacement observed in the mappings shown in Figure~\ref{fig:acqw_mappings}. The measured Z-contrast and displacement profiles are obtained by averaging the corresponding mappings for \hkl(110) planes along the growth direction . (a) composition profiles of elementary indium, aluminum and gallium from the sigmoidal model. (b) Measured and modeled Z-contrast from the compositions in (a). (c) Measured and modeled displacement along the growth direction using the compositions in (a) and assuming pseudomorphic accommodation on InP.}
    \label{fig:acqw_profiles}
\end{figure}

An atomically-resolved HAADF-STEM micrograph of the ACQW structure in sample A is shown in Figure~\ref{fig:acqw_mappings}(a). Following the algorithm in Reference~\cite{pantzas_experimental_2021},  mappings of the  Z-contrast and the in-plane $\left(\varepsilon_{110}\right)$ and out-of-plane $\left(\varepsilon_{001}\right)$ strain were computed from this micrograph and are shown in panels (b),(c), and (d), respectively. Line profiles where extracted from these mappings by averaging values along single atomic monolayers perpendicular to the  \hkl<001> growth direction. The line profiles for the Z-contrast and $\varepsilon_{001}$ are shown in red in the middle and bottom plot of Figure~\ref{fig:acqw_profiles}\footnote{The line profile for $\varepsilon_{110}$ is not shown, as the average strain is equal to zero within the precision of the algorithm, confirming that the ACQW structure is pseudomorphically accommodated on the InP substrate.}. The Z-contrast $\mathcal{R}$ was modeled using:
\begin{equation}
    \mathcal{R}(z) = \frac{1}{\mathrm{Z}_\textnormal{ref}}\Bigg(\sum_{m=0}^{2}\mathrm{Z}_m^{1.8}\mathcal{C}_m(z)+\mathrm{Z}_{3}^{1.8}\Bigg),
\end{equation}
where $\mathrm{Z}_m$ is atomic number of alloy element $m$ and $\mathcal{C}_m$ its composition from Equation~\ref{eq:sumNL}, with elements indium, aluminum, gallium and arsenic are numbered $m=0$ through 3. The reference is taken to be InGaAs lattice-matched to InP, i.e. $\mathrm{Z}_\textnormal{ref}=0.53\mathrm{Z}_{0}^{1.8}+0.47\mathrm{Z}_{2}^{1.8}+\mathrm{Z}_{3}^{1.8}$ \cite{pantzasSubnanometricallyResolvedChemical2016}.
For pseudomorphic accommodation on the InP substrate, $\varepsilon_{001}$ follows:
\begin{equation}
    \varepsilon_{001} (z)= \Bigg(1+\frac{C_{12}(z)}{C_{11}(z)} \Bigg)\frac{a_0(z)-a_\mathrm{InP}}{a_\mathrm{InP}},
    \label{eq:strain}
\end{equation}
where $C_{ij}(z)$ are the position-dependent, alloy elastic constants, $a_0(z)$ the strain-free alloy lattice parameter, and $a_\mathrm{InP}$ the lattice parameter of the InP substrate. The elastic constants and strain-free lattice parameters are linearly interpolated from the corresponding parameters of the end-point binaries and their respective compositions in the alloy from Equation~\ref{eq:sumNL}. The modeled curves that yield the best fit to experiment are shown in blue in the middle and bottom plot of Figure~\ref{fig:acqw_profiles}. To obtain the fits, a single value for all characteristic lengths $L_n$, equal to 1.5 atomic monolayers or, equivalently, \SI{2.2}{\nano\meter} and the set $\{c_{m,n}\}_{M,N}$  given in Table~\ref{tab:acqw_comp} were used. The ordinates $z_n$ of the interfaces were obtained from the first derivative of $\mathcal{R}$, that depends linearly on $\mathcal{C}_m$ and, hence, reaches an extremum at the same positions. The resulting models for the distribution of indium, aluminum, and gallium along the \hkl<001> direction for each element is shown in Figure~\ref{fig:acqw_profiles} (top). These modeled composition profiles show that the \SI{1.4}{\nano\meter} thick InAlAs barrier deviates significantly from a nominally lattice-matched InAlAs alloy, containing at most only \SI{25}{\percent} aluminum, and \SI{35}{\percent} gallium, and \SI{40}{\percent} indium. The subsequent \SI{2.4}{\nano\meter} thick InGaAs well also contains up to \SI{10}{\percent} aluminum. 

\begin{figure}
    \centering
    \includegraphics[width=0.7\textwidth]{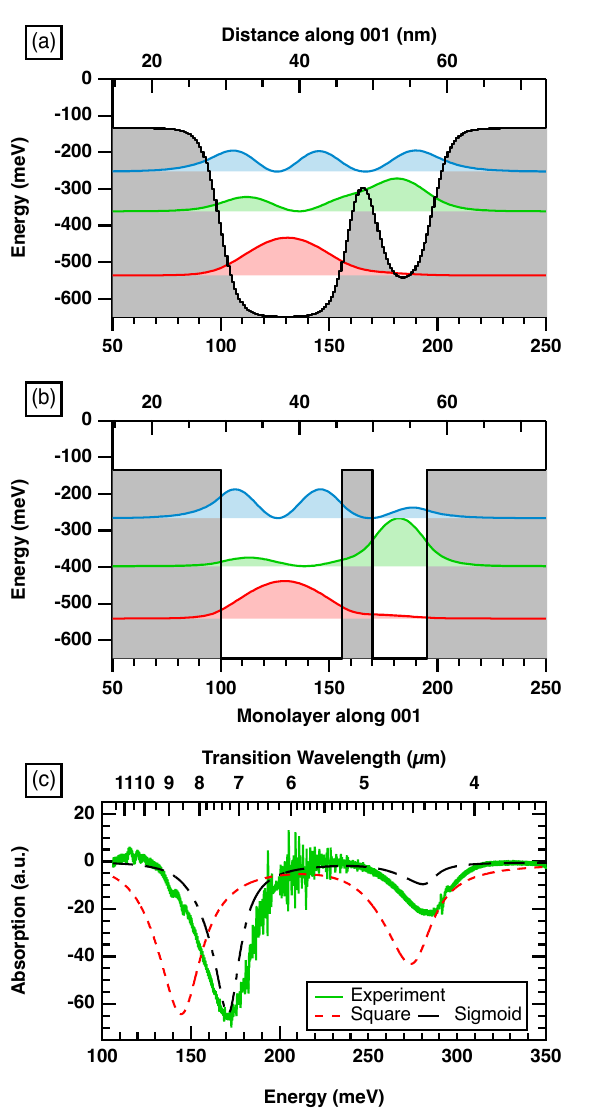}
    \caption{Band-structure simulations of the ACQW structure, carried out using (a) the sigmoidal profile described previously, and (b) ideally abrupt, perfectly nominal wells and barrier. (c) Experimental absorption spectrum (green, solid line) compared against the simulated absorption using the sigmoid model (black, dot-dashed) and the ideally abrupt profile (red, dashed).}
    \label{fig:acqw_bandstructure}
\end{figure}

The effect of these deviations from an ideally abrupt composition profile was evaluated by numerically computing the absorption for both the measured composition profile and an ideal square-band equivalent, and comparing them to experiment. The band structure computed of the measured composition profiles is shown in Figure~\ref{fig:acqw_bandstructure}~(a), and its square-band equivalent in Figure~\ref{fig:acqw_bandstructure}~(b). In both cases, levels $E_1$ and $E_2$ are coupled, with $E_1$ mostly localized in the \SI{5.6}{\nano\meter} thick well and $E_2$ mostly in the \SI{2.5}{\nano\meter} thick well. The third level, $E_3$ significantly differs in the two structures: in the actual structure level $E_3$ is above the barrier and distributed across the entire width of the ACQW structure, whereas it is mostly localized in the \SI{5.6}{\nano\meter} thick well in the case of the square-band potential. The differences in the band-structures are slight. They do, however, yield a significant difference in the absorption spectra shown Figure~\ref{fig:acqw_bandstructure}. The peak absorption for the $E_{12}$ transition appears at \SI{170}{\milli\electronvolt} in the experimental spectrum (green solid). This peak is exactly reproduced by the sigmoidal profile (black, dot-dashed), whereas the square-band equivalent $E_{12}$ is offset and appears \SI{144}{\milli\electronvolt}. The sigmoidal profile also more accurately reproduces the $E_{13}$ transition, predicting it \SI{278}{\milli\electronvolt} for an experimental peak at \SI{281}{\milli\electronvolt}, versus \SI{272}{\milli\electronvolt} for the square-band equivalent.

While the deviations of the square-band model from the experiment appear small, they are significant when designing unipolar devices emitting, modulating, or detecting in the mid-infrared. The difference between \SI{170}{\milli\electronvolt} and \SI{144}{\milli\electronvolt} in the $E_{12}$ transition of the ACQW corresponds to a \SI{1.4}{\micro\meter} shift in the operating wavelength. Furthermore, it is important to recall that the sample observed using HAADF-STEM is distinct from the one on which absorption measurements were carried out. This showcases that the methodology proposed here is in fact a calibration of the interfaces in a given reactor and for a given set of conditions, that remains valid for an extended amount of time. Therefore, the proposed empirical model and its demonstrated predictive strength can help one better design the target device and overcome limitations of epitaxial crystals. The authors have already used this information to produce a  Stark-effect modulator  used in Reference~\cite{dely10GbitFree2022}, that yielded a free-space communication mid-infrared channel at \SI{9}{\micro\meter} with a record speed of 10~Gbps. 

Beyond the demonstrated utility in terms of device design, the proposed composition model provides a useful metric for the analysis of epitaxial crystals and their interfaces. In the present contribution, it was applied to (In,Al,Ga)As alloys and explains how the flow modulation scheme introduced in Reference~\cite{pantzasSubnanometricallyResolvedChemical2016} helps partially counteract grading. Indeed, if one considers the derivative of the composition at a given interface and for a given alloy, one sees that it is proportional to $(c_{n+1}-c_n)/4L_n$. Hence, increasing the flow of aluminum at the barrier increases the slope by acting on $\Delta c = c_{n+1}-c_n$. In the authors experience, this effect is much larger than that of temperature, or growth stops on the characteristic interface length $L$. It would be interesting to apply  Equation~\ref{eq:sumNL} to a variety of other alloys, in particularly, those that tend to exhibit dissymetric interfaces, such as antimonide-arsenide alloys. The model could be further improved to reflect precursor flow ratios instead of the set of compositions $\{c_{m,n}\}_{M,N}$ and better understand the impact that mass-transport limited growth has on interfaces. Finally, additional data could be mined from the mappings in Figure~\ref{fig:acqw_mappings}, in particular the distribution of elements within a single atomic monolayer that can be used to more accurately model and predict the spectral linewidth of optical transitions \cite{ndebeka-bandouRelevanceIntraIntersubband2012}.

\begin{table}[]
    \centering
    \begin{tabular}{ccccccc}
    \toprule
        & $\mathbf{c_{m,n}}$ &  \textbf{0} & \textbf{1} & \textbf{2} & \textbf{3} & \textbf{4}\\

         \midrule
        \textbf{[In]} & \textbf{0}  & 0.522 & 0.532 & 0.45 & 0.5 & 0.522 \\
        \textbf{[Al]} & \textbf{1} & 0.478 & 0 & 0.53 & 0.08 & 0.478 \\
        \textbf{[Ga]} &\textbf{2} & 0 & 0.468 & 0.02 & 0.42 & 0 \\
        \bottomrule
    \end{tabular}
    \caption{Effective compositions $c_{m,n}$ used in the Equation~\ref{eq:sumNL} to model the compositions of each layer in the HAADF-STEM micrograph of the ACQW structure presented in Figure~\ref{fig:acqw_mappings}-(a). The best fit for the characteristic transition length was found to be \num[]{1.5} monolayers, or \SI{0.44}{\nano\meter}.}
    \label{tab:acqw_comp}
\end{table}

\section{Conclusion}\label{sec:conclusion}

A generalized empirical model for epitaxial semiconductor superlattices has been described. The model describes superlattices with an arbitrary number of interfaces and alloy elements. It accurately reflects finite grading at interfaces between wells and barriers, reproducing the microstructure of the crystal lattice as observed in atomically resolved HAADF-STEM images. The model is used to obtain the band structure of the superlattice and predict the absorption spectrum of an asymmetric coupled quantum-well structure, accurately reproducing the transitions observed in the experimental absorption spectrum of an asymmetric coupled quantum-well structure.

\begin{acknowledgements}
The authors would like to thankfully acknowledge funding from the French Renatech network, the ANR Labex TEMPOS (ANR-10-EQPX0050), the ANR project LIGNEDEMIR (ANR-18-CE09-0035), and European Union’s Horizon 2020 research and innovation programme under grant agreement No 828893 (cFLOW project).
\end{acknowledgements}

\bibliography{references}

\begin{thebibliography}{23}%
\makeatletter
\providecommand \@ifxundefined [1]{%
 \@ifx{#1\undefined}
}%
\providecommand \@ifnum [1]{%
 \ifnum #1\expandafter \@firstoftwo
 \else \expandafter \@secondoftwo
 \fi
}%
\providecommand \@ifx [1]{%
 \ifx #1\expandafter \@firstoftwo
 \else \expandafter \@secondoftwo
 \fi
}%
\providecommand \natexlab [1]{#1}%
\providecommand \enquote  [1]{``#1''}%
\providecommand \bibnamefont  [1]{#1}%
\providecommand \bibfnamefont [1]{#1}%
\providecommand \citenamefont [1]{#1}%
\providecommand \href@noop [0]{\@secondoftwo}%
\providecommand \href [0]{\begingroup \@sanitize@url \@href}%
\providecommand \@href[1]{\@@startlink{#1}\@@href}%
\providecommand \@@href[1]{\endgroup#1\@@endlink}%
\providecommand \@sanitize@url [0]{\catcode `\\12\catcode `\$12\catcode
  `\&12\catcode `\#12\catcode `\^12\catcode `\_12\catcode `\%12\relax}%
\providecommand \@@startlink[1]{}%
\providecommand \@@endlink[0]{}%
\providecommand \url  [0]{\begingroup\@sanitize@url \@url }%
\providecommand \@url [1]{\endgroup\@href {#1}{\urlprefix }}%
\providecommand \urlprefix  [0]{URL }%
\providecommand \Eprint [0]{\href }%
\providecommand \doibase [0]{https://doi.org/}%
\providecommand \selectlanguage [0]{\@gobble}%
\providecommand \bibinfo  [0]{\@secondoftwo}%
\providecommand \bibfield  [0]{\@secondoftwo}%
\providecommand \translation [1]{[#1]}%
\providecommand \BibitemOpen [0]{}%
\providecommand \bibitemStop [0]{}%
\providecommand \bibitemNoStop [0]{.\EOS\space}%
\providecommand \EOS [0]{\spacefactor3000\relax}%
\providecommand \BibitemShut  [1]{\csname bibitem#1\endcsname}%
\let\auto@bib@innerbib\@empty
\bibitem [{\citenamefont {Martini}\ \emph {et~al.}(2001)\citenamefont
  {Martini}, \citenamefont {Gmachl}, \citenamefont {Falciglia}, \citenamefont
  {Curti}, \citenamefont {Bethea}, \citenamefont {Capasso}, \citenamefont
  {Whittaker}, \citenamefont {Paiella}, \citenamefont {Tredicucci},
  \citenamefont {Hutchinson}, \citenamefont {Sivco},\ and\ \citenamefont
  {Cho}}]{martini_high-speed_2001}%
  \BibitemOpen
  \bibfield  {author} {\bibinfo {author} {\bibfnamefont {R.}~\bibnamefont
  {Martini}}, \bibinfo {author} {\bibfnamefont {C.}~\bibnamefont {Gmachl}},
  \bibinfo {author} {\bibfnamefont {J.}~\bibnamefont {Falciglia}}, \bibinfo
  {author} {\bibfnamefont {F.}~\bibnamefont {Curti}}, \bibinfo {author}
  {\bibfnamefont {C.}~\bibnamefont {Bethea}}, \bibinfo {author} {\bibfnamefont
  {F.}~\bibnamefont {Capasso}}, \bibinfo {author} {\bibfnamefont
  {E.}~\bibnamefont {Whittaker}}, \bibinfo {author} {\bibfnamefont
  {R.}~\bibnamefont {Paiella}}, \bibinfo {author} {\bibfnamefont
  {A.}~\bibnamefont {Tredicucci}}, \bibinfo {author} {\bibfnamefont
  {A.}~\bibnamefont {Hutchinson}}, \bibinfo {author} {\bibfnamefont
  {D.}~\bibnamefont {Sivco}},\ and\ \bibinfo {author} {\bibfnamefont
  {A.}~\bibnamefont {Cho}},\ }\bibfield  {title} {\bibinfo {title} {High-speed
  modulation and free-space optical audio/video transmission using quantum
  cascade lasers},\ }\href {https://doi.org/10.1049/el:20010102} {\bibfield
  {journal} {\bibinfo  {journal} {ELECTRONICS LETTERS}\ }\textbf {\bibinfo
  {volume} {37}},\ \bibinfo {pages} {191} (\bibinfo {year} {2001})}\BibitemShut
  {NoStop}%
\bibitem [{\citenamefont {Diaz}\ \emph {et~al.}(2016)\citenamefont {Diaz},
  \citenamefont {Thomas}, \citenamefont {Castillo}, \citenamefont {Gross},\
  and\ \citenamefont {Moshary}}]{diaz_active_2016}%
  \BibitemOpen
  \bibfield  {author} {\bibinfo {author} {\bibfnamefont {A.}~\bibnamefont
  {Diaz}}, \bibinfo {author} {\bibfnamefont {B.}~\bibnamefont {Thomas}},
  \bibinfo {author} {\bibfnamefont {P.}~\bibnamefont {Castillo}}, \bibinfo
  {author} {\bibfnamefont {B.}~\bibnamefont {Gross}},\ and\ \bibinfo {author}
  {\bibfnamefont {F.}~\bibnamefont {Moshary}},\ }\bibfield  {title} {\bibinfo
  {title} {Active standoff detection of {CH}¡sub¿4¡/sub¿ and
  {N}¡sub¿2¡/sub¿{O} leaks using hard-target backscattered light using an
  open-path quantum cascade laser sensor},\ }\bibfield  {journal} {\bibinfo
  {journal} {APPLIED PHYSICS B-LASERS AND OPTICS}\ }\textbf {\bibinfo {volume}
  {122}},\ \href {https://doi.org/10.1007/s00340-016-6396-x}
  {10.1007/s00340-016-6396-x} (\bibinfo {year} {2016})\BibitemShut {NoStop}%
\bibitem [{\citenamefont {Villares}\ \emph {et~al.}(2014)\citenamefont
  {Villares}, \citenamefont {Hugi}, \citenamefont {Blaser},\ and\ \citenamefont
  {Faist}}]{villares_dual-comb_2014}%
  \BibitemOpen
  \bibfield  {author} {\bibinfo {author} {\bibfnamefont {G.}~\bibnamefont
  {Villares}}, \bibinfo {author} {\bibfnamefont {A.}~\bibnamefont {Hugi}},
  \bibinfo {author} {\bibfnamefont {S.}~\bibnamefont {Blaser}},\ and\ \bibinfo
  {author} {\bibfnamefont {J.}~\bibnamefont {Faist}},\ }\bibfield  {title}
  {\bibinfo {title} {Dual-comb spectroscopy based on quantum-cascade-laser
  frequency combs},\ }\bibfield  {journal} {\bibinfo  {journal} {NATURE
  COMMUNICATIONS}\ }\textbf {\bibinfo {volume} {5}},\ \href
  {https://doi.org/10.1038/ncomms6192} {10.1038/ncomms6192} (\bibinfo {year}
  {2014})\BibitemShut {NoStop}%
\bibitem [{\citenamefont {Hale}\ \emph {et~al.}(2000)\citenamefont {Hale},
  \citenamefont {Bester}, \citenamefont {Danchi}, \citenamefont {Fitelson},
  \citenamefont {Hoss}, \citenamefont {Lipman}, \citenamefont {Monnier},
  \citenamefont {Tuthill},\ and\ \citenamefont {Townes}}]{hale_berkeley_2000}%
  \BibitemOpen
  \bibfield  {author} {\bibinfo {author} {\bibfnamefont {D.}~\bibnamefont
  {Hale}}, \bibinfo {author} {\bibfnamefont {M.}~\bibnamefont {Bester}},
  \bibinfo {author} {\bibfnamefont {W.}~\bibnamefont {Danchi}}, \bibinfo
  {author} {\bibfnamefont {W.}~\bibnamefont {Fitelson}}, \bibinfo {author}
  {\bibfnamefont {S.}~\bibnamefont {Hoss}}, \bibinfo {author} {\bibfnamefont
  {E.}~\bibnamefont {Lipman}}, \bibinfo {author} {\bibfnamefont
  {J.}~\bibnamefont {Monnier}}, \bibinfo {author} {\bibfnamefont
  {P.}~\bibnamefont {Tuthill}},\ and\ \bibinfo {author} {\bibfnamefont
  {C.}~\bibnamefont {Townes}},\ }\bibfield  {title} {\bibinfo {title} {The
  {Berkeley} infrared spatial interferometer: {A} heterodyne stellar
  interferometer for the mid-infrared},\ }\href
  {https://doi.org/10.1086/309049} {\bibfield  {journal} {\bibinfo  {journal}
  {ASTROPHYSICAL JOURNAL}\ }\textbf {\bibinfo {volume} {537}},\ \bibinfo
  {pages} {998} (\bibinfo {year} {2000})}\BibitemShut {NoStop}%
\bibitem [{\citenamefont {Dely}\ \emph {et~al.}(2022)\citenamefont {Dely},
  \citenamefont {Bonazzi}, \citenamefont {Spitz}, \citenamefont {Rodriguez},
  \citenamefont {Gacemi}, \citenamefont {Todorov}, \citenamefont {Pantzas},
  \citenamefont {Beaudoin}, \citenamefont {Sagnes}, \citenamefont {Li},
  \citenamefont {Davies}, \citenamefont {Linfield}, \citenamefont {Grillot},
  \citenamefont {Vasanelli},\ and\ \citenamefont
  {Sirtori}}]{dely10GbitFree2022}%
  \BibitemOpen
  \bibfield  {author} {\bibinfo {author} {\bibfnamefont {H.}~\bibnamefont
  {Dely}}, \bibinfo {author} {\bibfnamefont {T.}~\bibnamefont {Bonazzi}},
  \bibinfo {author} {\bibfnamefont {O.}~\bibnamefont {Spitz}}, \bibinfo
  {author} {\bibfnamefont {E.}~\bibnamefont {Rodriguez}}, \bibinfo {author}
  {\bibfnamefont {D.}~\bibnamefont {Gacemi}}, \bibinfo {author} {\bibfnamefont
  {Y.}~\bibnamefont {Todorov}}, \bibinfo {author} {\bibfnamefont
  {K.}~\bibnamefont {Pantzas}}, \bibinfo {author} {\bibfnamefont
  {G.}~\bibnamefont {Beaudoin}}, \bibinfo {author} {\bibfnamefont
  {I.}~\bibnamefont {Sagnes}}, \bibinfo {author} {\bibfnamefont
  {L.}~\bibnamefont {Li}}, \bibinfo {author} {\bibfnamefont {A.~G.}\
  \bibnamefont {Davies}}, \bibinfo {author} {\bibfnamefont {E.~H.}\
  \bibnamefont {Linfield}}, \bibinfo {author} {\bibfnamefont {F.}~\bibnamefont
  {Grillot}}, \bibinfo {author} {\bibfnamefont {A.}~\bibnamefont {Vasanelli}},\
  and\ \bibinfo {author} {\bibfnamefont {C.}~\bibnamefont {Sirtori}},\
  }\bibfield  {title} {\bibinfo {title} {10 {{Gbit}} s-1 {{Free Space Data
  Transmission}} at 9 {$M$}m {{Wavelength With Unipolar Quantum
  Optoelectronics}}},\ }\href {https://doi.org/10.1002/lpor.202100414}
  {\bibfield  {journal} {\bibinfo  {journal} {Laser \& Photonics Reviews}\
  }\textbf {\bibinfo {volume} {16}},\ \bibinfo {pages} {2100414} (\bibinfo
  {year} {2022})}\BibitemShut {NoStop}%
\bibitem [{\citenamefont {Saemian}\ \emph {et~al.}(2024)\citenamefont
  {Saemian}, \citenamefont {Balzo}, \citenamefont {Gacemi}, \citenamefont
  {Todorov}, \citenamefont {Rodriguez}, \citenamefont {Lopez}, \citenamefont
  {Darquié}, \citenamefont {Li}, \citenamefont {Davies}, \citenamefont
  {Linfield}, \citenamefont {Vasanelli},\ and\ \citenamefont
  {Sirtori}}]{saemian_ultra-sensitive_2024}%
  \BibitemOpen
  \bibfield  {author} {\bibinfo {author} {\bibfnamefont {M.}~\bibnamefont
  {Saemian}}, \bibinfo {author} {\bibfnamefont {L.~D.}\ \bibnamefont {Balzo}},
  \bibinfo {author} {\bibfnamefont {D.}~\bibnamefont {Gacemi}}, \bibinfo
  {author} {\bibfnamefont {Y.}~\bibnamefont {Todorov}}, \bibinfo {author}
  {\bibfnamefont {E.}~\bibnamefont {Rodriguez}}, \bibinfo {author}
  {\bibfnamefont {O.}~\bibnamefont {Lopez}}, \bibinfo {author} {\bibfnamefont
  {B.}~\bibnamefont {Darquié}}, \bibinfo {author} {\bibfnamefont
  {L.}~\bibnamefont {Li}}, \bibinfo {author} {\bibfnamefont {A.~G.}\
  \bibnamefont {Davies}}, \bibinfo {author} {\bibfnamefont {E.}~\bibnamefont
  {Linfield}}, \bibinfo {author} {\bibfnamefont {A.}~\bibnamefont
  {Vasanelli}},\ and\ \bibinfo {author} {\bibfnamefont {C.}~\bibnamefont
  {Sirtori}},\ }\bibfield  {title} {\bibinfo {title} {Ultra-sensitive
  heterodyne detection at room temperature in the atmospheric windows},\ }\href
  {https://doi.org/10.1515/nanoph-2023-0787} {\bibfield  {journal} {\bibinfo
  {journal} {Nanophotonics}\ }\textbf {\bibinfo {volume} {13}},\ \bibinfo
  {pages} {1765} (\bibinfo {year} {2024})}\BibitemShut {NoStop}%
\bibitem [{\citenamefont {Ashuach}\ \emph {et~al.}(2013)\citenamefont
  {Ashuach}, \citenamefont {Kauffmann}, \citenamefont {Saguy}, \citenamefont
  {Grossman}, \citenamefont {Klin}, \citenamefont {Weiss},\ and\ \citenamefont
  {Zolotoyabko}}]{ashuachQuantificationAtomicIntermixing2013}%
  \BibitemOpen
  \bibfield  {author} {\bibinfo {author} {\bibfnamefont {Y.}~\bibnamefont
  {Ashuach}}, \bibinfo {author} {\bibfnamefont {Y.}~\bibnamefont {Kauffmann}},
  \bibinfo {author} {\bibfnamefont {C.}~\bibnamefont {Saguy}}, \bibinfo
  {author} {\bibfnamefont {S.}~\bibnamefont {Grossman}}, \bibinfo {author}
  {\bibfnamefont {O.}~\bibnamefont {Klin}}, \bibinfo {author} {\bibfnamefont
  {E.}~\bibnamefont {Weiss}},\ and\ \bibinfo {author} {\bibfnamefont
  {E.}~\bibnamefont {Zolotoyabko}},\ }\bibfield  {title} {\bibinfo {title}
  {Quantification of atomic intermixing in short-period {{InAs}}/{{GaSb}}
  superlattices for infrared photodetectors},\ }\href
  {https://doi.org/10.1063/1.4804252} {\bibfield  {journal} {\bibinfo
  {journal} {Journal of Applied Physics}\ }\textbf {\bibinfo {volume} {113}},\
  \bibinfo {pages} {184305} (\bibinfo {year} {2013})}\BibitemShut {NoStop}%
\bibitem [{\citenamefont {Patil}\ \emph {et~al.}(2017)\citenamefont {Patil},
  \citenamefont {Luna}, \citenamefont {Matsuda}, \citenamefont {Yamada},
  \citenamefont {Kamiya}, \citenamefont {Ishikawa},\ and\ \citenamefont
  {Shimomura}}]{patilGaAsBiGaAsMultiquantum2017}%
  \BibitemOpen
  \bibfield  {author} {\bibinfo {author} {\bibfnamefont {P.~K.}\ \bibnamefont
  {Patil}}, \bibinfo {author} {\bibfnamefont {E.}~\bibnamefont {Luna}},
  \bibinfo {author} {\bibfnamefont {T.}~\bibnamefont {Matsuda}}, \bibinfo
  {author} {\bibfnamefont {K.}~\bibnamefont {Yamada}}, \bibinfo {author}
  {\bibfnamefont {K.}~\bibnamefont {Kamiya}}, \bibinfo {author} {\bibfnamefont
  {F.}~\bibnamefont {Ishikawa}},\ and\ \bibinfo {author} {\bibfnamefont
  {S.}~\bibnamefont {Shimomura}},\ }\bibfield  {title} {\bibinfo {title}
  {{{GaAsBi}}/{{GaAs}} multi-quantum well {{LED}} grown by molecular beam
  epitaxy using a two-substrate-temperature technique},\ }\href
  {https://doi.org/10.1088/1361-6528/aa596c} {\bibfield  {journal} {\bibinfo
  {journal} {Nanotechnology}\ }\textbf {\bibinfo {volume} {28}},\ \bibinfo
  {pages} {105702} (\bibinfo {year} {2017})}\BibitemShut {NoStop}%
\bibitem [{\citenamefont {Luna}\ \emph {et~al.}(2012)\citenamefont {Luna},
  \citenamefont {Guzm{\'a}n}, \citenamefont {Trampert},\ and\ \citenamefont
  {{\'A}lvarez}}]{lunaCriticalRoleTwoDimensional2012}%
  \BibitemOpen
  \bibfield  {author} {\bibinfo {author} {\bibfnamefont {E.}~\bibnamefont
  {Luna}}, \bibinfo {author} {\bibfnamefont {{\'A}.}~\bibnamefont
  {Guzm{\'a}n}}, \bibinfo {author} {\bibfnamefont {A.}~\bibnamefont
  {Trampert}},\ and\ \bibinfo {author} {\bibfnamefont {G.}~\bibnamefont
  {{\'A}lvarez}},\ }\bibfield  {title} {\bibinfo {title} {Critical {{Role}} of
  {{Two-Dimensional Island-Mediated Growth}} on the {{Formation}} of
  {{Semiconductor Heterointerfaces}}},\ }\href
  {https://doi.org/10.1103/PhysRevLett.109.126101} {\bibfield  {journal}
  {\bibinfo  {journal} {Physical Review Letters}\ }\textbf {\bibinfo {volume}
  {109}},\ \bibinfo {pages} {126101} (\bibinfo {year} {2012})}\BibitemShut
  {NoStop}%
\bibitem [{\citenamefont {Luna}\ \emph {et~al.}(2009)\citenamefont {Luna},
  \citenamefont {Ishikawa}, \citenamefont {Satpati}, \citenamefont {Rodriguez},
  \citenamefont {Tourni{\'e}},\ and\ \citenamefont
  {Trampert}}]{lunaInterfacePropertiesGa2009}%
  \BibitemOpen
  \bibfield  {author} {\bibinfo {author} {\bibfnamefont {E.}~\bibnamefont
  {Luna}}, \bibinfo {author} {\bibfnamefont {F.}~\bibnamefont {Ishikawa}},
  \bibinfo {author} {\bibfnamefont {B.}~\bibnamefont {Satpati}}, \bibinfo
  {author} {\bibfnamefont {J.~B.}\ \bibnamefont {Rodriguez}}, \bibinfo {author}
  {\bibfnamefont {E.}~\bibnamefont {Tourni{\'e}}},\ and\ \bibinfo {author}
  {\bibfnamefont {A.}~\bibnamefont {Trampert}},\ }\bibfield  {title} {\bibinfo
  {title} {Interface properties of ({{Ga}},{{In}})({{N}},{{As}}) and
  ({{Ga}},{{In}})({{As}},{{Sb}}) materials systems grown by molecular beam
  epitaxy},\ }\href {https://doi.org/10.1016/j.jcrysgro.2008.10.039} {\bibfield
   {journal} {\bibinfo  {journal} {Journal of Crystal Growth}\ }\bibinfo
  {series} {International {{Conference}} on {{Molecular Beam Epitaxy}}
  ({{MBE-XV}})},\ \textbf {\bibinfo {volume} {311}},\ \bibinfo {pages} {1739}
  (\bibinfo {year} {2009})}\BibitemShut {NoStop}%
\bibitem [{\citenamefont {Luna}\ \emph {et~al.}(2008)\citenamefont {Luna},
  \citenamefont {Ishikawa}, \citenamefont {Batista},\ and\ \citenamefont
  {Trampert}}]{lunaIndiumDistributionInterfaces2008}%
  \BibitemOpen
  \bibfield  {author} {\bibinfo {author} {\bibfnamefont {E.}~\bibnamefont
  {Luna}}, \bibinfo {author} {\bibfnamefont {F.}~\bibnamefont {Ishikawa}},
  \bibinfo {author} {\bibfnamefont {P.~D.}\ \bibnamefont {Batista}},\ and\
  \bibinfo {author} {\bibfnamefont {A.}~\bibnamefont {Trampert}},\ }\bibfield
  {title} {\bibinfo {title} {Indium distribution at the interfaces of
  ({{Ga}},{{In}})({{N}},{{As}})/{{GaAs}} quantum wells},\ }\href
  {https://doi.org/10.1063/1.2907508} {\bibfield  {journal} {\bibinfo
  {journal} {Applied Physics Letters}\ }\textbf {\bibinfo {volume} {92}},\
  \bibinfo {pages} {141913} (\bibinfo {year} {2008})}\BibitemShut {NoStop}%
\bibitem [{\citenamefont {Hulko}\ \emph {et~al.}(2008)\citenamefont {Hulko},
  \citenamefont {Thompson},\ and\ \citenamefont
  {Simmons}}]{hulkoComparisonQuantumWell2008}%
  \BibitemOpen
  \bibfield  {author} {\bibinfo {author} {\bibfnamefont {O.}~\bibnamefont
  {Hulko}}, \bibinfo {author} {\bibfnamefont {D.~A.}\ \bibnamefont
  {Thompson}},\ and\ \bibinfo {author} {\bibfnamefont {J.~G.}\ \bibnamefont
  {Simmons}},\ }\bibfield  {title} {\bibinfo {title} {Comparison of {{Quantum
  Well Interdiffusion}} on {{Group III}}, {{Group V}}, and {{Combined Groups
  III}} and {{V Sublattices}} in {{GaAs-Based Structures}}},\ }\href
  {https://doi.org/10.1109/JSTQE.2008.920041} {\bibfield  {journal} {\bibinfo
  {journal} {IEEE Journal of Selected Topics in Quantum Electronics}\ }\textbf
  {\bibinfo {volume} {14}},\ \bibinfo {pages} {1104} (\bibinfo {year}
  {2008})}\BibitemShut {NoStop}%
\bibitem [{\citenamefont {Prosa}\ \emph {et~al.}(2011)\citenamefont {Prosa},
  \citenamefont {Clifton}, \citenamefont {Zhong}, \citenamefont {Tyagi},
  \citenamefont {Shivaraman}, \citenamefont {DenBaars}, \citenamefont
  {Nakamura},\ and\ \citenamefont {Speck}}]{prosaAtomProbeAnalysis2011}%
  \BibitemOpen
  \bibfield  {author} {\bibinfo {author} {\bibfnamefont {T.~J.}\ \bibnamefont
  {Prosa}}, \bibinfo {author} {\bibfnamefont {P.~H.}\ \bibnamefont {Clifton}},
  \bibinfo {author} {\bibfnamefont {H.}~\bibnamefont {Zhong}}, \bibinfo
  {author} {\bibfnamefont {A.}~\bibnamefont {Tyagi}}, \bibinfo {author}
  {\bibfnamefont {R.}~\bibnamefont {Shivaraman}}, \bibinfo {author}
  {\bibfnamefont {S.~P.}\ \bibnamefont {DenBaars}}, \bibinfo {author}
  {\bibfnamefont {S.}~\bibnamefont {Nakamura}},\ and\ \bibinfo {author}
  {\bibfnamefont {J.~S.}\ \bibnamefont {Speck}},\ }\bibfield  {title} {\bibinfo
  {title} {Atom probe analysis of interfacial abruptness and clustering within
  a single {{InxGa1}}-{{xN}} quantum well device on semipolar
  (101\textasciimacron 1\textasciimacron ) {{GaN}} substrate},\ }\href
  {https://doi.org/10.1063/1.3589370} {\bibfield  {journal} {\bibinfo
  {journal} {Applied Physics Letters}\ }\textbf {\bibinfo {volume} {98}},\
  \bibinfo {pages} {191903} (\bibinfo {year} {2011})}\BibitemShut {NoStop}%
\bibitem [{\citenamefont {Grange}\ \emph {et~al.}(2020)\citenamefont {Grange},
  \citenamefont {Mukherjee}, \citenamefont {Capellini}, \citenamefont
  {Montanari}, \citenamefont {Persichetti}, \citenamefont {Di~Gaspare},
  \citenamefont {Birner}, \citenamefont {Attiaoui}, \citenamefont
  {Moutanabbir}, \citenamefont {Virgilio},\ and\ \citenamefont
  {De~Seta}}]{grangeAtomicScaleInsightsSemiconductor2020}%
  \BibitemOpen
  \bibfield  {author} {\bibinfo {author} {\bibfnamefont {T.}~\bibnamefont
  {Grange}}, \bibinfo {author} {\bibfnamefont {S.}~\bibnamefont {Mukherjee}},
  \bibinfo {author} {\bibfnamefont {G.}~\bibnamefont {Capellini}}, \bibinfo
  {author} {\bibfnamefont {M.}~\bibnamefont {Montanari}}, \bibinfo {author}
  {\bibfnamefont {L.}~\bibnamefont {Persichetti}}, \bibinfo {author}
  {\bibfnamefont {L.}~\bibnamefont {Di~Gaspare}}, \bibinfo {author}
  {\bibfnamefont {S.}~\bibnamefont {Birner}}, \bibinfo {author} {\bibfnamefont
  {A.}~\bibnamefont {Attiaoui}}, \bibinfo {author} {\bibfnamefont
  {O.}~\bibnamefont {Moutanabbir}}, \bibinfo {author} {\bibfnamefont
  {M.}~\bibnamefont {Virgilio}},\ and\ \bibinfo {author} {\bibfnamefont
  {M.}~\bibnamefont {De~Seta}},\ }\bibfield  {title} {\bibinfo {title}
  {Atomic-{{Scale Insights}} into {{Semiconductor Heterostructures}}: {{From
  Experimental Three-Dimensional Analysis}} of the {{Interface}} to a
  {{Generalized Theory}} of {{Interfacial Roughness Scattering}}},\ }\href
  {https://doi.org/10.1103/PhysRevApplied.13.044062} {\bibfield  {journal}
  {\bibinfo  {journal} {Physical Review Applied}\ }\textbf {\bibinfo {volume}
  {13}},\ \bibinfo {pages} {044062} (\bibinfo {year} {2020})}\BibitemShut
  {NoStop}%
\bibitem [{\citenamefont {M{\"u}ller}\ \emph {et~al.}(2012)\citenamefont
  {M{\"u}ller}, \citenamefont {Gault}, \citenamefont {Field}, \citenamefont
  {Sullivan}, \citenamefont {Smith},\ and\ \citenamefont
  {Grovenor}}]{mullerInterfacialChemistryInAs2012}%
  \BibitemOpen
  \bibfield  {author} {\bibinfo {author} {\bibfnamefont {M.}~\bibnamefont
  {M{\"u}ller}}, \bibinfo {author} {\bibfnamefont {B.}~\bibnamefont {Gault}},
  \bibinfo {author} {\bibfnamefont {M.}~\bibnamefont {Field}}, \bibinfo
  {author} {\bibfnamefont {G.~J.}\ \bibnamefont {Sullivan}}, \bibinfo {author}
  {\bibfnamefont {G.~D.~W.}\ \bibnamefont {Smith}},\ and\ \bibinfo {author}
  {\bibfnamefont {C.~R.~M.}\ \bibnamefont {Grovenor}},\ }\bibfield  {title}
  {\bibinfo {title} {Interfacial chemistry in an {{InAs}}/{{GaSb}} superlattice
  studied by pulsed laser atom probe tomography},\ }\href
  {https://doi.org/10.1063/1.3688045} {\bibfield  {journal} {\bibinfo
  {journal} {Applied Physics Letters}\ }\textbf {\bibinfo {volume} {100}},\
  \bibinfo {pages} {083109} (\bibinfo {year} {2012})}\BibitemShut {NoStop}%
\bibitem [{\citenamefont {Wang}\ \emph {et~al.}(2017)\citenamefont {Wang},
  \citenamefont {Schwarz}, \citenamefont {Siriani}, \citenamefont {Connors},
  \citenamefont {Missaggia}, \citenamefont {Calawa}, \citenamefont {McNulty},
  \citenamefont {Akey}, \citenamefont {Zheng}, \citenamefont {Donnelly},
  \citenamefont {Mansuripur},\ and\ \citenamefont
  {Capasso}}]{wangSensitivityHeterointerfacesEmission2017}%
  \BibitemOpen
  \bibfield  {author} {\bibinfo {author} {\bibfnamefont {C.~A.}\ \bibnamefont
  {Wang}}, \bibinfo {author} {\bibfnamefont {B.}~\bibnamefont {Schwarz}},
  \bibinfo {author} {\bibfnamefont {D.~F.}\ \bibnamefont {Siriani}}, \bibinfo
  {author} {\bibfnamefont {M.~K.}\ \bibnamefont {Connors}}, \bibinfo {author}
  {\bibfnamefont {L.~J.}\ \bibnamefont {Missaggia}}, \bibinfo {author}
  {\bibfnamefont {D.~R.}\ \bibnamefont {Calawa}}, \bibinfo {author}
  {\bibfnamefont {D.}~\bibnamefont {McNulty}}, \bibinfo {author} {\bibfnamefont
  {A.}~\bibnamefont {Akey}}, \bibinfo {author} {\bibfnamefont {M.~C.}\
  \bibnamefont {Zheng}}, \bibinfo {author} {\bibfnamefont {J.~P.}\ \bibnamefont
  {Donnelly}}, \bibinfo {author} {\bibfnamefont {T.~S.}\ \bibnamefont
  {Mansuripur}},\ and\ \bibinfo {author} {\bibfnamefont {F.}~\bibnamefont
  {Capasso}},\ }\bibfield  {title} {\bibinfo {title} {Sensitivity of
  heterointerfaces on emission wavelength of quantum cascade lasers},\ }\href
  {https://doi.org/10.1016/j.jcrysgro.2016.11.029} {\bibfield  {journal}
  {\bibinfo  {journal} {Journal of Crystal Growth}\ }\bibinfo {series}
  {Proceedings of the 18th {{International Conference}} on {{Metal Organic
  Vapor Phase Epitaxy}}},\ \textbf {\bibinfo {volume} {464}},\ \bibinfo {pages}
  {215} (\bibinfo {year} {2017})}\BibitemShut {NoStop}%
\bibitem [{\citenamefont {Luna}\ \emph {et~al.}(2010)\citenamefont {Luna},
  \citenamefont {Satpati}, \citenamefont {Rodriguez}, \citenamefont {Baranov},
  \citenamefont {Tourni{\'e}},\ and\ \citenamefont
  {Trampert}}]{lunaInterfacialIntermixingInAs2010}%
  \BibitemOpen
  \bibfield  {author} {\bibinfo {author} {\bibfnamefont {E.}~\bibnamefont
  {Luna}}, \bibinfo {author} {\bibfnamefont {B.}~\bibnamefont {Satpati}},
  \bibinfo {author} {\bibfnamefont {J.~B.}\ \bibnamefont {Rodriguez}}, \bibinfo
  {author} {\bibfnamefont {A.~N.}\ \bibnamefont {Baranov}}, \bibinfo {author}
  {\bibfnamefont {E.}~\bibnamefont {Tourni{\'e}}},\ and\ \bibinfo {author}
  {\bibfnamefont {A.}~\bibnamefont {Trampert}},\ }\bibfield  {title} {\bibinfo
  {title} {Interfacial intermixing in {{InAs}}/{{GaSb}}
  short-period-superlattices grown by molecular beam epitaxy},\ }\href
  {https://doi.org/10.1063/1.3291666} {\bibfield  {journal} {\bibinfo
  {journal} {Applied Physics Letters}\ }\textbf {\bibinfo {volume} {96}},\
  \bibinfo {pages} {021904} (\bibinfo {year} {2010})}\BibitemShut {NoStop}%
\bibitem [{\citenamefont {Pantzas}\ \emph {et~al.}(2016)\citenamefont
  {Pantzas}, \citenamefont {Beaudoin}, \citenamefont {Patriarche},
  \citenamefont {Largeau}, \citenamefont {Mauguin}, \citenamefont {Pegolotti},
  \citenamefont {Vasanelli}, \citenamefont {Calvar}, \citenamefont {Amanti},
  \citenamefont {Sirtori},\ and\ \citenamefont
  {Sagnes}}]{pantzasSubnanometricallyResolvedChemical2016}%
  \BibitemOpen
  \bibfield  {author} {\bibinfo {author} {\bibfnamefont {K.}~\bibnamefont
  {Pantzas}}, \bibinfo {author} {\bibfnamefont {G.}~\bibnamefont {Beaudoin}},
  \bibinfo {author} {\bibfnamefont {G.}~\bibnamefont {Patriarche}}, \bibinfo
  {author} {\bibfnamefont {L.}~\bibnamefont {Largeau}}, \bibinfo {author}
  {\bibfnamefont {O.}~\bibnamefont {Mauguin}}, \bibinfo {author} {\bibfnamefont
  {G.}~\bibnamefont {Pegolotti}}, \bibinfo {author} {\bibfnamefont
  {A.}~\bibnamefont {Vasanelli}}, \bibinfo {author} {\bibfnamefont
  {A.}~\bibnamefont {Calvar}}, \bibinfo {author} {\bibfnamefont
  {M.}~\bibnamefont {Amanti}}, \bibinfo {author} {\bibfnamefont
  {C.}~\bibnamefont {Sirtori}},\ and\ \bibinfo {author} {\bibfnamefont
  {I.}~\bibnamefont {Sagnes}},\ }\bibfield  {title} {\bibinfo {title}
  {Sub-nanometrically resolved chemical mappings of quantum-cascade laser
  active regions},\ }\href@noop {} {\bibfield  {journal} {\bibinfo  {journal}
  {Semiconductor Science and Technology}\ }\textbf {\bibinfo {volume} {31}}
  (\bibinfo {year} {2016})}\BibitemShut {NoStop}%
\bibitem [{\citenamefont {Pantzas}\ and\ \citenamefont
  {Patriarche}(2021{\natexlab{a}})}]{pantzasExperimentalQuantificationAtomicallyresolved2021}%
  \BibitemOpen
  \bibfield  {author} {\bibinfo {author} {\bibfnamefont {K.}~\bibnamefont
  {Pantzas}}\ and\ \bibinfo {author} {\bibfnamefont {G.}~\bibnamefont
  {Patriarche}},\ }\bibfield  {title} {\bibinfo {title} {Experimental
  quantification of atomically-resolved {{HAADF-STEM}} images using {{EDX}}},\
  }\href@noop {} {\bibfield  {journal} {\bibinfo  {journal} {Ultramicroscopy}\
  }\textbf {\bibinfo {volume} {220}},\ \bibinfo {pages} {113152} (\bibinfo
  {year} {2021}{\natexlab{a}})}\BibitemShut {NoStop}%
\bibitem [{\citenamefont {Sirtori}\ \emph {et~al.}(1994)\citenamefont
  {Sirtori}, \citenamefont {Capasso}, \citenamefont {Faist},\ and\
  \citenamefont {Scandolo}}]{Sirtori1994}%
  \BibitemOpen
  \bibfield  {author} {\bibinfo {author} {\bibfnamefont {C.}~\bibnamefont
  {Sirtori}}, \bibinfo {author} {\bibfnamefont {F.}~\bibnamefont {Capasso}},
  \bibinfo {author} {\bibfnamefont {J.}~\bibnamefont {Faist}},\ and\ \bibinfo
  {author} {\bibfnamefont {S.}~\bibnamefont {Scandolo}},\ }\bibfield  {title}
  {\bibinfo {title} {Nonparabolicity and a sum rule associated with
  bound-to-bound and bound-to-continuum intersubband transitions in quantum
  wells},\ }\href {https://doi.org/10.1103/PhysRevB.50.8663} {\bibfield
  {journal} {\bibinfo  {journal} {Phys. Rev. B}\ }\textbf {\bibinfo {volume}
  {50}},\ \bibinfo {pages} {8663} (\bibinfo {year} {1994})}\BibitemShut
  {NoStop}%
\bibitem [{\citenamefont {Terazzi}(2012)}]{terazzi_transport_2012}%
  \BibitemOpen
  \bibfield  {author} {\bibinfo {author} {\bibfnamefont {R.~L.}\ \bibnamefont
  {Terazzi}},\ }\emph {\bibinfo {title} {Transport in quantum cascade
  lasers}},\ \href {https://doi.org/10.3929/ethz-a-007057728} {\bibinfo {type}
  {Doctoral {Thesis}}},\ \bibinfo  {school} {ETH Zurich}, \bibinfo {address}
  {Zürich} (\bibinfo {year} {2012})\BibitemShut {NoStop}%
\bibitem [{\citenamefont {Pantzas}\ and\ \citenamefont
  {Patriarche}(2021{\natexlab{b}})}]{pantzas_experimental_2021}%
  \BibitemOpen
  \bibfield  {author} {\bibinfo {author} {\bibfnamefont {K.}~\bibnamefont
  {Pantzas}}\ and\ \bibinfo {author} {\bibfnamefont {G.}~\bibnamefont
  {Patriarche}},\ }\bibfield  {title} {\bibinfo {title} {Experimental
  quantification of atomically-resolved {HAADF}-{STEM} images using {EDX}},\
  }\href@noop {} {\bibfield  {journal} {\bibinfo  {journal} {Ultramicroscopy}\
  }\textbf {\bibinfo {volume} {220}},\ \bibinfo {pages} {113152} (\bibinfo
  {year} {2021}{\natexlab{b}})}\BibitemShut {NoStop}%
\bibitem [{\citenamefont {{Ndebeka-Bandou}}\ \emph {et~al.}(2012)\citenamefont
  {{Ndebeka-Bandou}}, \citenamefont {Carosella}, \citenamefont {Ferreira},
  \citenamefont {Wacker},\ and\ \citenamefont
  {Bastard}}]{ndebeka-bandouRelevanceIntraIntersubband2012}%
  \BibitemOpen
  \bibfield  {author} {\bibinfo {author} {\bibfnamefont {C.}~\bibnamefont
  {{Ndebeka-Bandou}}}, \bibinfo {author} {\bibfnamefont {F.}~\bibnamefont
  {Carosella}}, \bibinfo {author} {\bibfnamefont {R.}~\bibnamefont {Ferreira}},
  \bibinfo {author} {\bibfnamefont {A.}~\bibnamefont {Wacker}},\ and\ \bibinfo
  {author} {\bibfnamefont {G.}~\bibnamefont {Bastard}},\ }\bibfield  {title}
  {\bibinfo {title} {Relevance of intra- and inter-subband scattering on the
  absorption in heterostructures},\ }\href {https://doi.org/10.1063/1.4766192}
  {\bibfield  {journal} {\bibinfo  {journal} {Applied Physics Letters}\
  }\textbf {\bibinfo {volume} {101}},\ \bibinfo {pages} {191104} (\bibinfo
  {year} {2012})}\BibitemShut {NoStop}%
\end{thebibliography}%

\end{document}